\documentclass[12pt,preprint]{aastex}
\usepackage{amsmath}
\usepackage{graphics}      % standard graphics specifications
\usepackage{graphicx}

\begin{document}
\newcommand{\kms}{\mbox{km~s$^{-1}$}}
\newcommand{\s}{\mbox{$''$}}
\newcommand{\mloss}{\mbox{$\dot{M}$}}
\newcommand{\my}{\mbox{$M_{\odot}$~yr$^{-1}$}}
\newcommand{\ls}{\mbox{$L_{\odot}$}}
\newcommand{\ms}{\mbox{$M_{\odot}$}}
\newcommand\mdot{$\dot{M}  $}

\title{Strong Variable Ultraviolet Emission from Y\,Gem: Accretion Activity in an AGB Star with a Binary Companion?}
\author{Raghvendra Sahai\altaffilmark{1}, James D. Neill\altaffilmark{2}, Armando Gil de Paz\altaffilmark{3},
 \& Carmen S{\'a}nchez Contreras\altaffilmark{4} 
}
%\email{sahai@jpl.nasa.gov}
\altaffiltext{1}{Jet Propulsion Laboratory, MS\,183-900, California Institute of Technology, Pasadena, CA 91109}
\altaffiltext{2}{California Institute of Technology, 1200 E. California Blvd.  MC278-17, Pasadena, CA, 91125, USA}
\altaffiltext{3}{Dpto. de Astrof\'{\i}sica, Universidad Complutense de Madrid, Madrid 28040, Spain}
\altaffiltext{4}{Astrobiology Center (CSIC-INTA), ESAC campus, E-28691 Villanueva de la Canada, Madrid, Spain}

\begin{abstract}
Binarity is believed to dramatically affect the history and geometry
of mass loss in AGB and post-AGB stars, but observational evidence of binarity is sorely lacking. As part of a project to look
for hot binary companions to cool AGB stars using the GALEX archive, we have discovered a late-M star, Y\,Gem, to be a source of
strong and variable UV emission. Y\,Gem is a prime example of the success of our technique of UV imaging of AGB stars in order to
search for binary companions. Y\,Gem's large and variable UV flux makes it one of the most prominent examples of a late AGB star
with a mass accreting binary companion. The UV emission is most likely due to emission associated with accretion
activity and a disk around a main-sequence companion star. The physical mechanism generating the UV emission is extremely
energetic,
with an integrated luminosity of a few$\times$\ls at its peak. We also find weak CO J=2-1 emission from Y\,Gem with a very narrow
line profile 
(FWHM of 3.4\,\kms). Such a narrow line is unlikely to arise in an outflow, and is consistent with emission from an orbiting,
molecular reservoir of radius 300\,AU. Y\,Gem may be the progenitor of the class of post-AGB stars which are binaries and possess
disks but no outflows. 
\end{abstract}


\section{Introduction}\label{intro}
Binarity, very common amongst pre-main-sequence stars (e.g., Bodenheimer et al. 2000), is widely believed to
dramatically affect the history and geometry of mass loss in late-AGB and post-AGB stars (e.g., Balick \& Frank 2002).
The evolutionary transition from the AGB to the planetary nebula (PN) phase is accompanied by significant changes in
the morphology of these objects -- the roughly round circumstellar mass-loss envelopes (CSEs) of AGB stars evolve into
pre-planetary-nebulae (PPNs) and PNs with a dazzling variety of shapes and intriguing aspherical symmetries (e.g$.$
Schwarz et al$.$ 1992, Sahai \& Trauger 1998, Sahai et al. 2007, Sahai et al. 2011). Binarity provides a source of
angular momentum, as
well as a preferred axis to a stellar system. Critical reviews (e.g., Soker 1998) of the properties of bipolar PNs
lead to the conclusion that binary models can explain all these properties, whereas
single-star models have many difficulties.
%(e.g$.$ Corradi \& Schwarz 1995) 

However, direct observational evidence for binarity in dying intermediate-mass stars is difficult to obtain. AGB stars are very
luminous ($\sim$\,10$^3$-10$^4$\,\ls) and surrounded by dusty envelopes, whereas nearby stellar companions are generally likely
to be significantly less luminous main-sequence stars or white dwarfs. Indirect techniques such as radial-velocity (RV)
measurements or photometric-variability (PV) measurements cannot be used for AGB stars since their strong periodic or irregular
variability intrinsic to their pulsating atmospheres masks the corresponding variability due to a companion. PV measurements of
the central stars of PNs (CSPNs) have resulted in detecting a total of about 40 binaries (e.g., Bond 2000, Miszalski et al.
2011), implying a 10-15\% fraction of detectable close binaries among randomly selected PNs. RV measurements of CSPNs and
post-AGB (pAGB) stars with prominent outflows (i.e., PPNs) have had very limited success (e.g., de Marco 2009, Hrivnak et al.
2011) because of the unknown contribution from stellar variability. Only for disk-prominent pAGB stars (dpAGB: Sahai et al.
2011), RV techniques have reached a 100\% success rate in detecting binarity (e.g., Van Winckel et al. 2009).

%* In the intro, second paragraph, I would differentiate much better between binary searches in PNe, post-AGB and AGB. The
%techniques used are namely very different: For PNe in the references covered, they are based on photometric techniques which are
%only sensitive to spiraled-in narrow systems. The radial velocity searches in PNe have proven to be difficult (see indeed the
%reference on de Marco (2004). Radial velocity programs on post-AGB stars are also relevant here and these are not covered: e.g.
%Hrivnak et al., 2011 (which show the lack of binary indications for post-AGB stars with a shell) and Van Winckel et al., 2009
%(100 percent binary rate for post-AGB stars with a disc) are relevant in this context. I suggest that the authors restructure
%this paragraph to make this difference clearer.

Thus new techniques to search for the presence of binary companions in AGB stars are of fundamental importance.
Currently, ultraviolet photometry appears to be the most promising technique 
of being able to discover a substantial number of binary AGB stars. An as yet untested method which could become important  
is to look for X-rays from AGB companions which have been spun-up via accretion (Soker \& Kastner 2002).
Most AGB stars with substantial mass-loss are relatively cool ($T_{eff}\lesssim3000$K) objects (spectral types $\sim$M6
or later), whereas any stellar companions and/or accretion disks around them are likely significantly hotter
($T_{eff}\gtrsim6000$K). 
Observed and model spectra of AGB stars show that their fluxes die rapidly at wavelengths
shortwards of about 2800\AA, hence significantly favorable secondary/accretion disk-to-primary flux contrast ratios
($\gtrsim$10) are reached in the NUV window of GALEX for companions of spectral type hotter
than about G0 ($T_{eff}$=6000\,K) and luminosity, $L\gtrsim1$\ls. We used UV imaging with GALEX to carry out a pilot
search for binary companions in AGB stars (Sahai et al. 2008; hereafter Setal08). 
In this survey, we detected all of our 21 targets in the GALEX NUV filter with signal-to-noise ratios (SNR)
$\gtrsim$20; 9 sources were detected in the FUV band with SNR$\gtrsim8\sigma$. In this Letter, we report the discovery
of UV emission from the AGB star, Y\,Gem, from archival GALEX data.

\section{Observations \& Results} Y\,Gem was in the field-of-view of the GALEX Medium Imaging Survey (MIS) and All-Sky Imaging
Survey (AIS) during 3 epochs from Jan 2006 to Jan 2008 that are roughly spaced apart by 1 year. We discovered Y\,Gem's
remarkable UV emission serendipitously while examining the GALEX archive for UV-emitting AGB stars. Y\,Gem, otherwise an
unremarkable, semi-regular pulsating star of spectral type M8, turns out to be the brightest UV-emitting object amongst our
sample of about one hundred M4--M8 stars with detected FUV emission (Sahai et al., in preparation). Following this discovery,
we searched for CO J=2-1 line
emission from Y\,Gem on 2011 February 11 and 15, using the ARO 10-m SMT on Mt.
Graham, AZ, with the 1\,mm dual-polarization receiver
employing ALMA Band 6 SBS mixers. Typical system temperatures were 
170-250\,K. The beam size was $\theta_b=32$\arcsec, and pointing accuracy is estimated to be about $\pm$4\arcsec.
Observations were conducted in beam-switching mode with
a $\pm$3\arcmin subreflector throw.

%epoch  NUV      FUV (microJy) FUV/NUV NUVtime  FUVtime
%AIS1   11040.4  15944.5       1.44
%AIS2   11153.8   5170.3       0.46
%<AIS>  11097.1  10557.4       0.95
%MIS     1613.2   1384.97      0.86    1531.05  1531.05

Y\,Gem was detected in both UV bands in each epoch, but its ultraviolet flux decreased dramatically from 2006 to 2008 (Table\,1)
-- i.e., by
a factor 12 (5.6) in the FUV(NUV) (Figure\,1a). A model atmosphere corresponding to an M\,8III spectral type from Lejeune et al.
(1997), fitted to the spectral-energy-distribution (SED) from blue to the far-infrared shows that the UV flux densities are much
larger than expected from Y\,Gem's photosphere (Figure\,1b). Several of the flux values in 2006 \& 2007 are large enough that
non-linearity
corrections, as estimated from Morrissey et al. (2007), were significant and were used to correct the former; the corrected
values are also provided in Table\,1. An examination of the time-tagged photon lists (averaged over 10\,s
bins), showed no significant variability of the UV flux densities within each epoch, i.e., over periods of about 110\,s during
Epoch 1 and 2, and 1531\,s during Epoch 3. The FUV flux appears to decrease much faster than the NUV, with the
ratio of the FUV-to-NUV wavelength-specific fluxes, $R_{FUV/NUV}$, taking uncorrected (corrected) values of 5.53 (4.54), 1.38
(1.47), and 2.15 (2.15) in 2006, 2007 and 2008, respectively.
Y\,Gem had a FUV flux of 16\,mJy in 2006, much above the range of fluxes, $\sim0.01-0.1$\,mJy, observed in the 9/21
sources in our pilot survey with FUV detections.
%FUV-to-NUV fluxes (in mJy) taking values of 1.44, 0.46 and 0.86 in 2006, 2007 and 2008, respectively.

The value of $R_{FUV/NUV}$ can be used to estimate the equivalent blackbody temperature of the emitting region, $T_{bb}$. 
%(about 3 at $T_{bb}=5\times10^4$\,K: Robinson et al. 2003). 
Since it is difficult to assess the relative line and continuum contributions to
the FUV and NUV fluxes, we follow Robinson et al.'s (2005) arguments for arbitrarily (but conservatively) assuming that half
(all) of the FUV (NUV) flux is continuum emission. Thus, we find that in 2006, $T_{bb}=(3-3.8)\times10^4$\,K (for the
uncorrected-to-corrected range of $R_{FUV/NUV}$ values) when the UV flux was at its peak, and $T_{bb}=1.7\times10^4$\,K, when the
UV flux was at its lowest. The physical mechanism generating the UV emission is extremely energetic, with an integrated luminosity
of $(2.4-4)$\ls~at its peak in 2006.

We detected weak CO J=2-1 emission from Y\,Gem (Figure\,2). The line profile appears centrally peaked, with a full-width at zero
intensity (FWZI) of 5\,\kms and an
integrated flux of 0.044\,K\,kms (1.36\,Jy\,\kms assuming an unresolved source).
The CO line-width (FWZI) in Y\,Gem is quite small, compared to typical values for dusty outflows in AGB stars,  $>$20\
\kms (e.g., CO catalog by Loup
et al. 1993). Although we cannot rule out the possibility that Y\,Gem's CO emission comes from an anomalously slow outflow, it is
more likely that it arises in a large, orbiting reservoir of molecular material in this object. Jura \& Kahane (1999) argue that
such narrow lines, also seen in a few other
evolved stars (e.g., EU\,And), most likely arise from long-lived reservoirs of orbiting gas of size (100--1000)\,AU. The dpAGB
star, AC\,Her, a known binary with a compact disk (Gielen et al. 2007), also shows a similarly narrow CO line (Alcolea \&
Bujarrabal 1991). 

%* The authors refer to similar narrow CO line profiles as found in J-type silicate stars by Jura et al., 1999. However, J-type
%silicate stars are very poorly understood (no orbits found so far, strange abundances are not understood etcc). In the context
%of the circumbinary small discs in the discussion, it is probably also relevant to refer to also old papers by Bujarrabal et
%al., 1988, A\&A 206, L17 and Alcolea \& Bujarrabal, 1991, A\&A 245, 499) in which single dish observations show very narrow CO 
% profiles in stars with a compact disc.

We estimate a distance D=580\,pc to Y\,Gem from its TMSS K-band magnitude $m_K=1.22$, following Kahane \& Jura (1994) who assume
that late-M semi-regular stars have absolute magnitudes $M_K=-7.6$.
%(Kahane \& Jura 1994, A\&A, 290, 183)(Fluxes were taken from the IRC catalog: TMSS). 
Assuming that the gas resides in a Keplerian disk around a star of 1\ms, the speed of 1.7\,\kms (assumed to be half the FWHM of
the CO line) indicates an orbital radius of 300\,AU. At this distance from the star (with luminosity of $5800$\ls), a blackbody
grain would be at a temperature of $T_d=140$\,K. We adopt this temperature for the circumstellar gas, following Jura \& Kahane's
(1999) similar assumption for the disk in EU\,And. Assuming optically-thin emission, the total gas mass is 
$M_g=2\times10^{-5}(4\times10^{-4}/f_{CO})$\ms, where $f_{CO}$ is the fractional CO-to-H$_2$ abundance. Given this low gas mass, 
the fact that we can discern no dust emission in Y\,Gem's SED -- the mid- and far-infared fluxes are well fitted with
photospheric emission from the M\,8 primary (Figure\,1b) -- is not surprising. Assuming a typical circumstellar gas-to-dust
ratio for AGB stars of 200, the expected 60 and 90\,\micron~fluxes from 140\,K dust are 0.5 and 0.1\,Jy, significantly smaller
than the photospheric flux at these wavelengths.

\section{Discussion}
Noting that the observed FUV fluxes in their sample were at least a factor $\sim$30 higher than that can be produced by the
continuum of the cool primary AGB star, Setal08 presented two plausible models for the FUV excesses -- (1) the excess is due to
the photosphere of a hot companion star, and (2) the excess is due to line and/or continuum emission {\it associated} with  
accretion activity and a disk around a companion star -- the accretion shock may reside on (a) the disk or (b) the stellar surface
of
the companion (as, e.g., in T Tauri stars: Calvet \& Gullbring 1998). A 3rd model is emission from a companion related to either
(a) flare activity on an M dwarf (dM) companion, or (b) coronal (magnetic) activity of a late-type (most likely $\sim$F7-M4)
main-sequence companion which has been spun-up by wind accretion from the primary AGB star (Soker \& Kastner 2002). A 4th
model is variable chromospheric-type line emission from the AGB star as has been inferred in the C star, TW\,Hor, based on its
IUE spectra (Querci \& Querci 1985). A 5th model follows from the proposal by Soker \& Kastner (2003) of the possibility of
long-duration flares in single AGB stars due to magnetic
reconnection events leading to X-ray emission. The only AGB star with an observed
X-ray transient is Mira A which is in a (symbiotic) binary system (Karovska et al. 2005); and a sensitive search for X-rays in two
(apparently) single AGB stars yielded
upper limits at levels less than (1--10)\% of expectations (Kastner \& Soker 2004. Hence it is not clear whether magnetic
reconnection events in single AGB stars can produce X-ray flares; however, it would be worth investigating whether the latter can
pwer the kind of strong, variable UV emission seen in Y\,Gem. 

%* Discussion: I is a bit confusing to read so a more structural discussion should be made. It is better to suggest all possible
%interpretations of the high UV flux in the first paragraph (so including what the authors call 4). As it is now, the list of
%possible mechanisms is given up to 3(a,b,c) and even more alternatives are added later in the text.
%* Discussion: the list of possible physical mechanisms is given with annotation 1,2, 3(a,b,c) and 2A, 2B, SK03 etc. this reads
%quite confusing.

We think model 3a is unlikely. Flares on M dwarfs have been seen
in the UV using GALEX (Welsh et al. 2007), however the decay time-scales for the flare emission tends to be rather short -- e.g.,
Fig.\,2 of Welsh et al. (2007) shows that most M-dwarf flare sources decay to half their peak flux in $\lesssim$100\,s. In
contrast, Y\,Gem's flux showed no significant variations during any of the 3 epochs. Furthermore, dM flares are significantly
less luminous in the NUV and FUV than Y\,Gem: e.g., scaled to Y\,Gem's estimated distance of 580\,pc (see below), the peak NUV
flux of the most luminous flare object (J$023955.52-072855.4$) out of a total of 49 dM stars in Welsh et al's survey (who only
report NUV fluxes) is roughly a factor 35 fainter than Y\,Gem's peak (corrected) NUV flux, and the giant flare in the dMe star
GJ\,3685A is a factor 80 fainter in the FUV than Y\,Gem (Robinson et al. 2005). 

Model 3b was proposed by Soker \& Kastner (2002) in order to explain the X-ray emission observed from the central binary stars of
PNs and one can imagine extrapolating it to predict variable chromospheric UV emission from the spun-up companion as well.
However, we do not think models 3b and 4 are plausible models for Y\,Gem because for
chromospheric emission, the ratio of the FUV-to-NUV fluxes (in mJy units) as measured in the GALEX bands, $r_{FUV/NUV}$, is much
smaller than unity (about 0.1--0.05 in TW\,Hor, and about 0.1 in the red supergiant $\alpha$\,Ori), and results from the presence
of strong emission lines in the NUV band. The large value of $r_{FUV/NUV}$ in Y\,Gem cannot be a result of dust attenuation
because this would either not significantly change the FUV-NUV color, or would redden the UV spectrum. 

%Welsh et al. find short-term NUV variations -- light curves

The highly variable UV flux in Y\,Gem supports model 2 above for this source; in this scenario, the UV emission from Y\,Gem may
indicate episodic accretion (a) onto the disk, from matter ejected by the AGB star (e.g., Mastrodemos \& Morris 1998), or (b)
onto the companion from the inner disk region. A likely example of
the type $2a$ mechanis is provided by Mira, in which the compact companion 
(separation $\sim$60\,AU) accretes matter in a disk from the primary's wind. Mira shows variable NUV and FUV line emission: UV
lines observed with IUE during 1979-95 faded by
a factor $>$20 by 1999-2001, and then started increasing back to their original levels by 2004 (Wood \& Karovska 2006);
the continuum was also variable. These authors concluded that the UV variability results from variations in the
accretion rate (Mira\,A contributes to the UV flux only at wavelengths $\gtrsim$2600\AA~and only near its
maximum). Mira was observed with GALEX on two different epochs separated by 3 years, and we find that its FUV and NUV fluxes 
changed dramatically over this period as well (Table\,1).

Noting that Y\,Gem has never been classified as a symbiotic star, it is
likely that its companion is a main-sequence star. If the emission is of type $2b$, this inference is supported by our estimate of
the emitting region's area, $A_{emm}$, which is $3.3\times10^{-3}$ and
$(2.1-3.3)\times10^{-3}$ times the total surface area of a Sun-sized companion in 2008 and 2006, respectively (based on its UV
flux and inferred blackbody temperature) -- it exceeds the surface
area of
a white dwarf by at least a factor 5. 
We note that, in published (magnetospheric) accretion-shock models for T\,Tauri stars
(Gullbring et al. 2000, Calvet \& Gullbring 1998), even the largest values of the energy flux of the accretion flow,
$\mathcal{F}\sim10^{12}$erg\,cm$^{-2}$\,s$^{-1}$ (which determines the spectral shape of the UV emission), cannot produce the
lowest value of $R_{FUV/NUV}$ seen in Y\,Gem. Thus, if Y\,Gem's UV emission is due to a magnetospheric accretion-shock on the
companion, then $\mathcal{F}$, and thus the mass accretion rate -- typically $(0.1-few)\times10^{-7}$\ms\,in these models -- has
to be significantly higher than in T\,Tauri stars.

%* Is there some literature with optical spectral information of Y Gem ? I would guess that Balmer line emission originating in
%the accretion disc could be expected.

Y\,Gem's relatively blue color over the wavelength region spanning the GALEX NUV--FUV bands is very similar to that of
the carbon star V\,Hya, in which Setal08 found $r_{FUV/NUV}\sim1.1$ in two separate epochs. V\,Hya is probably the
best example to-date of an evolved star with an active, collimated outflow,
dense equatorially-flattened structures possibly related to a central accretion disk, and an
inferred binary companion and/or accretion disk from UV excesses (Setal08)
%, Sahai et al. 2003,\,2010; Hirano et al. 2004).

%Considering also the star's fast rotation revealed by the photospheric lines, we conclude that V Hya is probably
%experiencing the short binary common envelope evolution phase between the AGB and the planetary nebula stage
% \bibitem[Kahane et al.(1996)]{1996A&A...314..871K} 
%   Kahane, C., Audinos, P., Barnbaum, C., \& Morris, M.\ 1996, \aap, 314, 871

%using Chandra (X-ray) and HST (optical) images, authors concluded that the outburst was likely associated 
% with the cool primary, Mira A, and could be a localized, long-duration flare near the stellar surface due 
%to a magnetic reconnection event, as proposed by Soker \& Kastner (2003).

Y\,Gem appears to be one of the most prominent examples of a late AGB star with a (inferred) mass-accreting binary companion, but
is not
alone, given the results of our survey (Setal08). However, being the brightest, it is the best one to begin investigating in
order to ultimately determine whether the kind of binary interaction observed in Mira is generalized among UV-variable
AGB stars -- such binary interactions are widely believed to be the key to the formation of aspherical structure observed so
commonly in post-AGB (pAGB) objects. The aspherical structure has been characterized via specific descriptors in a comprehensive 
morphological classification system, developed for PPNs and extended to young PNs (Sahai et al. 2007,\,2011). The descriptors
primarily cover the shapes of the extended nebula and the central region. Two major classes of pAGB objects show
equatorially-flattened central structures. The first class of pAGB objects consists of PPNs and PNs; a large fraction of these
harbor overdense, dusty equatorial waists that are quite large; e.g., in several PPNs, the
waists can be seen to display sharp outer (radial) edges, with radii typically $\gtrsim 1000$\,AU (Sahai et al. 2007). In the
second 
class, namely dpAGB objects, there is strong evidence for medium-sized ($\sim$50\,AU) disks (e.g., van Winckel et al. 2008) and
little or no 
nebulosity. The relatively small CO line-width in Y\,Gem, implying the presence of a gravitationally bound
structure, rather than a typical AGB dusty outflow, suggests that it will most likely evolve into a
dpAGB object, rather than a PPN. 

Ultraviolet spectroscopic
observations of Y\,Gem using HST's Cosmic Origins Spectrograph are needed to probe the
nature of the UV emission -- i.e., if it is composed solely of emission lines arising in an accretion disk, or if there
is a substantial continuum contribution (e.g., from the disk itself or the companion's photosphere). Any emission lines
detected can provide constraints on the physical parameters of the accretion disk such as the (electron) density,
emission measure, and thus the emitting volume (e.g., Reimers \& Cassatella 1985). 
Multi-epoch observations (either at UV or radio wavelengths) can constrain the size of the emitting region depending on the
time-scale of the variability. In  
a type $2b$ mechanism, one might expect gyroresonance or
gyrosynchrotron emission at radio wavelengths from Y\,Gem -- e.g., Skinner \& Brown (1994) conclude that a likely model for the
2--3.6\,cm
emission in the pre-main-sequence star, T\,Tau S, is non-thermal gyrosynchrotron emission arising in a scaled-up solar-like
flare. Observations of high-J CO lines such as J=6-5, 7-6, 8-7, and 9-8 with the Herschel Space
Observatory are necessary to determine the temperature of the molecular gas in the disk.

\acknowledgements
We thank the staff of the Arizona Radio Observatory for granting us observing time. We thank Noam Soker and Joel Kastner for their
valuable comments on an earlier version of this paper. RS's contribution to the
research described here was carried out at the Jet Propulsion Laboratory, California Institute of Technology, under a
contract with NASA. Financial support was provided by NASA through a Long Term Space Astrophysics and GALEX GO award.

%\clearpage
\begin{figure}[hbt]
\resizebox{1.0\textwidth}{!}{\includegraphics{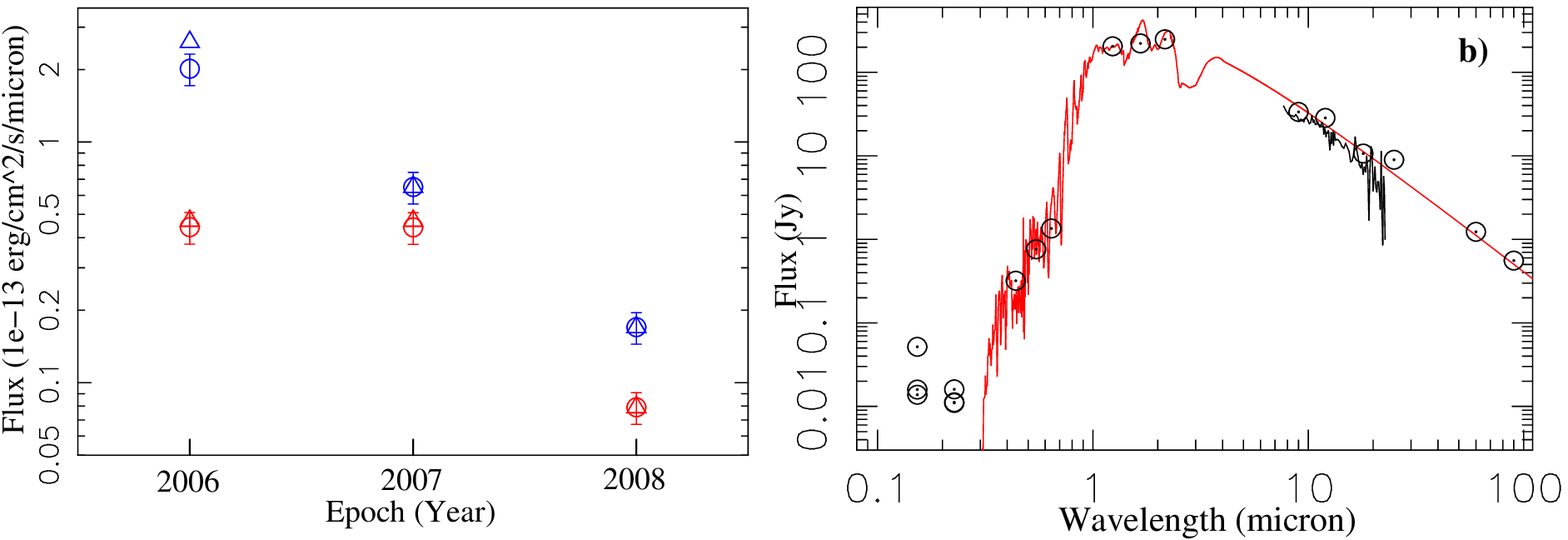}}
{\caption{(a) The variable UV emission from Y\,Gem. Both 
uncorrected values (circles) and corrected ones (for non-linearity) (triangles) of the FUV (blue symbols) \& NUV (red symbols)
fluxes vs. epoch are shown (the 0.15-mag
%(uncorrected for non-linearity)
error bars on the uncorrected values are the maximum zero point error that the
GALEX images have had since the early mission data was released). 
(b) The observed SED: photometry from GALEX (UV), USNO-B.1 and GSC2.3.2 (B,V,R), 
2MASS (J,H,K), AKARI-MID (9, 18\micron), AKARI-FIS (90\,\micron), 
IRAS (12, 25, 60\,\micron), is shown as ({\it black circles}), and the IRAS/LRS
spectrum as a {\it black curve}. A low-gravity T$_{eff}=2800$\,K model atmosphere spectrum from Lejeune et al. (1997),
corresponding to Y\,Gem's M\,8III spectral type ({\it red curve}) has been fitted to the SED.}
}
\label{flximg}
\end{figure}

\begin{figure}[hbt]
\resizebox{0.75\textwidth}{!}{\includegraphics{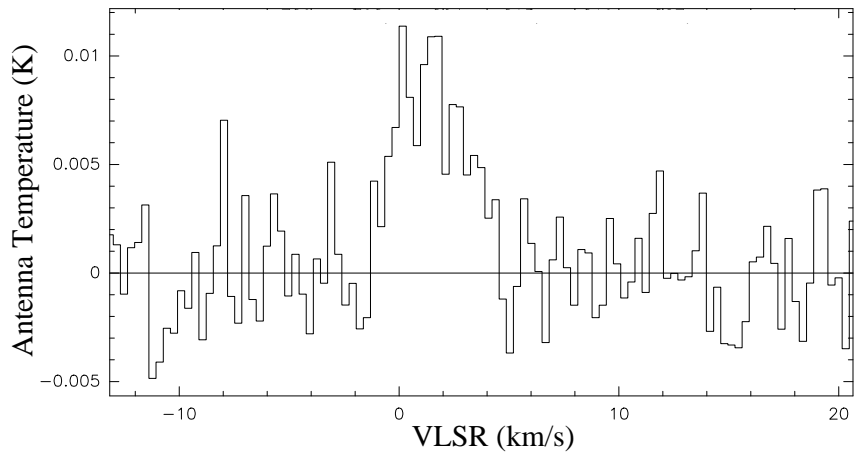}}
{\caption{CO J=2-1 line emission from Y\,Gem, observed with the Arizona Radio Observatory's 10-m SMT mm-wave
telescope. The temperature scale is measured as $T_A^*$, where the source's radiation
temperature is defined as $T_R=T_A^*$/$\eta_b$, and $\eta_b$ (=0.74) is the main-beam efficiency.}
}
\label{co21}
\end{figure}
% SMT data, spectrum, fig file in /u4/sahai/data/galex/agb/Y_Gem

\begin{table}
\caption{GALEX UV Photometry of Y\,Gem \& Mira}
\begin{tabular}{l c c c c c c c c c}\hline\noalign{\smallskip}
Source  & Epoch & Time & FUV   & NUV & FUV$_c$ & NUV$_c$ & FUV$_c$ & NUV$_c$ & (FUV/NUV)$_c$\\
 \multicolumn{1}{c}{} & 
 \multicolumn{1}{c}{yyyy-mm} &
 \multicolumn{1}{c}{sec} &
 \multicolumn{4}{c}{cps} &
 \multicolumn{2}{c}{10$^{-13}$erg\,s$^{-1}$\,cm$^{-2}$\,\AA$^{-1}$} \\
%        & yyyy-mm & sec  & cps &     & cps    &        & 10$^{-13}$erg\,s$^{-1}$\,cm$^{-2}$\,\AA$^{-1}$ & \\
\hline\noalign{\smallskip}
Y\,Gem  & 2006-01  & 114   & 143.4& 214.3 & 185  & 227   & 2.6  & 0.47 & 5.53\\
Y\,Gem  & 2007-02  & 109   & 46.5 & 216.5 & 46.5 & 230   & 0.65 & 0.47 & 1.38\\
Y\,Gem  & 2008-01  & 1531    & 12.4 & 38.1  & 12.4 & 38.1  & 0.17 & 0.079 & 2.15 \\
\hline\noalign{\smallskip}
Mira    & 2003-11  & 109   & 28.1 & 731.4 & 28.1 & 1032  & 0.39 & 2.13  &  0.18 \\
Mira    & 2006-11  & 11324   & 3.56 & 49.2  & 3.56 & 49.2  & 0.05 & 0.10 & 0.5 \\
\hline\noalign{\smallskip}
\end{tabular}
\end{table}
% if exp times differ by less than 5% of average time, report average time

\end{document}